\RequirePackage{fix-cm}
\documentclass[smallextended]{svjour3}       
\smartqed  

\usepackage{graphicx}

\usepackage{subfigure}
\usepackage{amssymb}
\usepackage{amsfonts, amsmath}
\usepackage{enumerate}
\usepackage{booktabs}
\usepackage{multirow}
\usepackage{changepage}
\usepackage{pdflscape}
\usepackage{multicol}
\usepackage[nolists]{endfloat}
\usepackage{url}
\usepackage{color}
\usepackage{latexsym}
\usepackage{amssymb}
\usepackage{color,soul}
\usepackage{natbib}
\begin{document}

\title{A proportional hazards model for interval-censored data subject to instantaneous failures
}


\author{Prabhashi W. Withana Gamage         \and
        Monica Chaudari   \and
        Christopher S. McMahan   \and
         Michael R. Kosorok
}


\institute{Prabhashi W. Withana Gamage \at
              Department of Mathematical Sciences, Clemson University, Clemson, SC 29634, U.S.A. \\
           \and
           Monica Chaudari \at
              Department of Biostatistics, University of North Carolina at Chapel Hill, Chapel Hill, NC 27599, U.S.A.\\
              \and
              Christopher S. McMahan \at
              Department of Mathematical Sciences, Clemson University, Clemson, SC 29634, U.S.A. \\
              \email{mcmaha2@clemson.edu}\\
              \and
              Michael R. Kosorok \at
              Department of Biostatistics, University of North Carolina at Chapel Hill, Chapel Hill, NC 27599, USA
}

\date{Received: date / Accepted: date}

\maketitle

\begin{abstract}
The proportional hazards (PH) model is arguably one of the most popular models used to analyze time to event data arising from clinical trials and longitudinal studies, among many others. In many such studies, the event time of interest is not directly observed but is known relative to periodic examination times; i.e., practitioners observe either current status or interval-censored data. The analysis of data of this structure is often fraught with many difficulties. Further exacerbating this issue, in some such studies the observed data also consists of instantaneous failures; i.e., the event times for several study units coincide exactly with the time at which the study begins. In light of these difficulties, this work focuses on developing a mixture model, under the PH assumptions, which can be used to analyze interval-censored data subject to instantaneous failures. To allow for modeling flexibility, two methods of estimating the unknown cumulative baseline hazard function are proposed; a fully parametric and a monotone spline representation are considered. Through a novel data augmentation procedure involving latent Poisson random variables, an expectation-maximization (EM) algorithm was developed to complete model fitting. The resulting EM algorithm is easy to implement and is computationally efficient. Moreover, through extensive simulation studies the proposed approach is shown to provide both reliable estimation and inference. 
\end{abstract}

\section{Introduction}
\label{intro}
Interval-censored data commonly arise in many clinical trials and longitudinal studies, and is characterized by the fact that the event time of interest is not directly observable, but rather is known relative to observation times. As a special case, current status data (or case-1 interval censoring) arise when there exists exactly one observation time per study unit; i.e., at the observation time one discovers whether or not the event of interest has occurred. Data of this structure often occurs in resource limited environments or due to destructive testing. Alternatively, general interval-censored data (or case-2 interval censoring) arise when multiple observation times are available for each study unit, and the event time can be ascertained relative to two observation times. It is well known that ignoring the structure of interval-censored data during an analysis can lead to biased estimation and inaccurate inference; see \cite{Odell1992}, \cite{Dorey1993}. Further exasperating this issue, some studies are subject to the occurrence of instantaneous failures; i.e., the event time of interest for a number of the study units occurs at time zero. This feature can occur as an artifact of the study design or may arise during an intent-to-treat analysis \citep{LiuYang2016, MatsuzakiNagatoshi2005, LambornYung2008}. For example, \cite{ChenLai2015} describes a registry based study of end-stage renal disease patients, with the time of enrollment corresponding to the time at which the patient first received dialysis. In this study, several of the patient expire during the first dialysis treatment, leading to the occurrence of an instantaneous failure. Similarly, \cite{LiemGraaf1997} describes an intent-to-treat clinical trial comparing conventional anterior surgery and laparoscopic surgery for repairing inguinal hernia.  In this study, various patients not receiving the allocated intervention, were inadequately excluded from the analysis to overcome issues such as consent withdrawal, procedure misfit etc. that would rightly attribute to instantaneous failures.
Survival data with instantaneous events is not uncommon in epidemiological and clinical studies, and for this reason, herein a general methodology under the proportional hazards (PH) model is developed for the analysis of interval-censored data subject to instantaneous failures.

Originally proposed by \cite{Cox1972}, the PH model has (arguably) become one of the most popular regression models for analyzing time-to-event data. For analyzing interval-censored data under the PH model, several notable contributions have been made in the recent years; e.g., see \cite{Finkelstein1986}, \cite{Satten1996}, \cite{GogginsFinkelstein}, \cite{GroeneboomWellner}, \cite{Pan1999}, \cite{Pan2000}, \cite{Goeteghebeur2000}, \cite{BetenskyLindsey2002}, \cite{CaiBetensky2003}, \cite{ZhangHua2010}, \cite{Sun2006}, \cite{zhangz2010}, and \cite{LiMa2013}. More recently, \cite{wang2016} developed a methodology under the PH model which can be used to accurately and reliably analyze interval-censored data. In particular, this approach makes use of a monotone spline representation to approximate the cumulative baseline hazard function. In doing so, an expectation-maximization (EM) algorithm is developed through a data augmentation scheme involving latent Poisson random variables which can be used to complete model fitting. It is worthwhile to note, that none of the aforementioned techniques were designed to account for the effects associated with instantaneous failures.

The phenomenon of instantaneous (or early) failures occur in many lifetime experiments; to include, but not limited to, reliability studies and clinical trials. In reliability studies, instantaneous failures may be attributable to inferior quality or faulty manufacturing, where as in clinical trials these events may manifest due to adverse reactions to treatments or clinical definitions of outcomes. When the failure times are exactly observed, as is the case in reliability studies, it is common to incorporate instantaneous failures through a mixture of parametric models, with one being degenerate at time zero; e.g., see \cite{Muralidharan1999}, \cite{KaleMuralidharan2000}, \cite{MurthyXie2004}, \cite{MuralidharanLathika2006}, \cite{Pham2007}, and \cite{Knopik2011}. In the case of interval-censored data, more common among epidemiological studies and clinical trials, accounting for instantaneous failures becomes a more tenuous task, with practically no guidance available among the existing literature. Arguably, in the context of interval-censored data, one could account for instantaneous failures by introducing an arbitrarily small constant for each as an observation time, and subsequently treat the instantaneous failures as left-censored observations. In doing so, methods for interval-censored data, such as those discussed above, could be employed. While this approach may seem enticing, in the case of a relatively large number of instantaneous failures it has several pitfalls. In particular, through numerical studies (results not shown) it has been determined that this approach when used in conjunction with modeling techniques such as those proposed in \cite{Pan1999} and \cite{wang2016} may lead to inaccurate estimation of the survival curves and/or the covariate effects. Further, after an extensive literature review, it does not appear that any methodology has previously been developed to specifically address data of this structure. For these reasons, herein a general methodology under the PH model is developed for the analysis of interval-censored data subject to instantaneous failures.

For the analysis of interval-censored data subject to instantaneous failures a new mixture model is proposed, which is a generalization of the semi-parametric PH model studied in \cite{wang2016}. The proposed PH model is developed under the standard PH assumption; i.e., the covariates provide for a multiplicative effect on the baseline risk of both experiencing a failure at time zero and thereafter. Two separate techniques are developed for the purposes of estimating the cumulative baseline hazard function. The first allows a practitioner to specify a parametric form (up to a collection of unknown coefficients) for the unknown function, while the second provides for more modeling flexibility through the use of the monotone splines of \cite{ramsay}. Under either formulation, a two-stage data augmentation scheme involving latent Poisson random variables is used to develop an efficient EM algorithm which can be used to estimate all of the unknown parameters. Through extensive simulation studies the proposed methodology is shown to provide reliable estimation and inference with respect to the covariate effects, baseline cumulative hazard function, and baseline probability of experiencing an instantaneous failure. 

The remainder of this article is organized as follows. Section 2 presents the development of the proposed model, the derivation of the EM algorithm, and outlines uncertainty quantification. The finite sample performance of the proposed approach is evaluated through extensive numerical studies, the features and results of which are provided in Section 3. 
Further, code which implements the proposed methodology has been added to the existing R software package \texttt{ICsurv} and is freely available from the CRAN (i.e., http://cran.us.rproject.org/).

\section{Model and Methodology}
Let $T$ denote the failure time of interest. Under the PH model, the survival function can be generally written as
\begin{equation}\label{S}
S(t|\mathbf{x})=S_0(t)^{e^{\mathbf{x}'\boldsymbol{\beta}}}
\end{equation}
where $\mathbf{x}$ is a $(r\times 1)$-dimensional vector of covariates, $\boldsymbol{\beta}$ is the corresponding vector of regression coefficients, and $S_{0}(t)$ is the baseline survival function. Under the phenomenon of interest, there is a baseline risk (probability) of experiencing an instantaneous failure; i.e., $S(0|\mathbf{x}=\mathbf{0}_r)=S_{0}(0)=1-p$, where $p\in[0,1]$ is the baseline risk and $\mathbf{0}_r$ is a $(r\times 1)$-dimensional vector of zeros. Thus, under the PH assumptions, the probability of experiencing an instantaneous failure, given the covariate information contained in $\mathbf{x}$, can be ascertained from \eqref{S} as
\begin{eqnarray*}
P(T=0|\mathbf{x}) & = & 1-S(0|\mathbf{x})\\
                  & = & 1-(1-p)^{e^{\mathbf{x}'\boldsymbol{\beta}}}.
\end{eqnarray*}
Similarly, given that an instantaneous failure does not occur, it is assumed that the failure time conditionally follows the standard PH model; i.e.,  
\begin{eqnarray*}
P(T>t|\mathbf{x},T>0)  =  1-F(t|\mathbf{x}),
\end{eqnarray*}
where $F(t|\mathbf{x})=1-\exp\{-\Lambda_{0}(t)\exp(\mathbf{x}'\boldsymbol{\beta})\}$ and $\Lambda_{0}(\cdot)$ is the usual cumulative baseline hazard function. Note, in order
for $F(\cdot|\mathbf{x})$ to be a proper cumulative distribution function, $\Lambda_{0}(\cdot)$ should be a monotone increasing function with $\Lambda_{0}(0)=0$. Thus, through an application of the Law of Total Probability, one has that 
\begin{eqnarray*}
P(T>t|\mathbf{x}) & = & P(T>t|\mathbf{x},T>0) P(T>0|\mathbf{x}) \\
                  & = & \{1-F(t|\mathbf{x})\}(1-p)^{e^{\mathbf{x}'\boldsymbol{\beta}}},
\end{eqnarray*}
for $t>0$.
Based on these assumptions, the cumulative distribution function of $T$ can be expressed as the following mixture model,
\[
   H(t|\mathbf{x})=\left\{
                \begin{array}{ll}
                  1-e^{-\alpha e^{\mathbf{x}'\boldsymbol{\beta}}}, &\textrm{ for } t=0,\\
                  1-e^{-\alpha e^{\mathbf{x}'\boldsymbol{\beta}}}\{1-F(t|\mathbf{x})\}, &\textrm{ for } t>0,
                 
                \end{array}
              \right.
  \]
where, for reasons that will shortly become apparent, $1-p$ is re-parametrized as $\exp(-\alpha)$, for $\alpha>0$.

\subsection{Observed data likelihood}
In scenarios where interval-censored data arise, one has that the failure time ($T$) is not directly observed, but rather is known relative to two observation times, say $L<R$; i.e., one has that $L<T<R$. In general, the four different outcomes considered here can be represented through the values of $L$ and $R$; i.e., an instantaneous failure ($L=R=0$), $T$ is left-censored ($0=L<R<\infty$), $T$ is interval-censored ($0<L<R<\infty$), and $T$ is right-censored ($0<L<R=\infty$). For notational convenience, let $\psi$ be an indicator denoting the event that $T$ is not an instantaneous failure, and $\delta_{1}$, $\delta_{2}$, and $\delta_{3}$ be censoring indicators denoting left-, interval-, and right-censoring, respectively; i.e., $\psi=I(T>0)$, $\delta_{1}=I(0=L<R<\infty)$, $\delta_{2}=I(0<L<R<\infty)$, and $\delta_{3}=(0<L<R=\infty)$.

In order to derive the observed data likelihood, it is assumed throughout that the individuals are independent, and that conditional on the covariates, the failure time for an individual is independent of the observational process. This assumption is common among the survival literature; see, e.g., \cite{zhangz2010}, \cite{LiuShen2009} and the references therein. The observed data collected on $n$ individuals is given by $\mathcal{D}=\{(L_{i}, R_{i}, {\bf x}_i, \psi_i, \delta_{i1}, \delta_{i2}, \delta_{i3}) ;  i=1,2,\dots,n\}$, which constitutes $n$ independent realization of $\{(L, R, {\bf x}, \psi, \delta_{1}, \delta_{2}, \delta_{3})$. Thus, under the aforementioned assumptions, the observed data likelihood is given by
\begin{equation}\label{eqn:odl1}
L_{obs}(\boldsymbol{\theta})=\prod_{i=1}^{n}\Big[F(R_i|{\bf x}_i)^{\delta_{i1}}\{F(R_i|{\bf x}_i)-F(L_i|{\bf x}_i)\}^{\delta_{i2}}\{1-F(L_i|{\bf x}_i)\}^{\delta_{i3}}\Big]^{\psi_i} \Big\{e^{-\alpha e^{\mathbf{x}'\boldsymbol{\beta}}}\Big\}^{\psi_i}\Big\{1-e^{-\alpha e^{\mathbf{x}'\boldsymbol{\beta}}}\Big\}^{1-\psi_i},
\end{equation}
where $\boldsymbol{\theta}$ represents the set of unknown parameters which are to be estimated. 

\subsection{Representations of $\Lambda_{0}(\cdot)$}
The unknown parameters in the observed likelihood involve the regression parameters $\boldsymbol{\beta}$, $\alpha$, and the cumulative baseline hazard function $\Lambda_{0}(\cdot)$. Herein, two techniques for modeling the cumulative baseline hazard function are discussed. The first approach considers the use of a fully parametric model, which is known up to a set of unknown coefficients. For example, a linear, quadratic, or logarithmic parametric model can be specified by setting $\Lambda_{0}(t)=\gamma_1t$, $\Lambda_{0}(t)=\gamma_1t+\gamma_2t^2$, and $\Lambda_{0}(t)=\gamma_1\log(1+t)$, respectively. Note, all of these models obey the constraints placed on $\Lambda_{0}(\cdot)$, as long as the $\gamma_l>0$, for $l=1,2$. In general, a parametric form for the cumulative baseline hazard model can be specified as 
\begin{equation}\label{eqn:spline}
\Lambda_{0}(t)=\sum_{l=1}^{k}\gamma_{l}b_{l}(t),
\end{equation}
where $b_l(\cdot)$ is a monotone increasing function, $b_l(0)=0$, and $\gamma_l>0$, for $l=1,...,k$. Under these mild conditions, it is easily verified that $\Lambda_{0}(\cdot)$ inherits the same properties, and therefore adheres to the aforementioned constraints.

The second approach, which is inspired by the works of \cite{wang2016}, \cite{LinWang2010}, \cite{WangDunson2011}, \cite{cai2011}, \cite{McMahanWangTebbs2013}, and \cite{WangWangMcMahan}, views $\Lambda_{0}(\cdot)$ as an unknown function, and hence an infinite dimensional parameter. To reduce the dimensionality of the problem, the monotone splines of \cite{ramsay} are used to approximate $\Lambda_{0}(\cdot)$. Structurally, this representation is identical to that of \eqref{eqn:spline} with the exception that $b_{l}(\cdot)$ is a spline basis function and $\gamma_{l}$ is an unknown spline coefficient, for $l=1,\dots ,k$. Again, it is required that $\gamma_{l}>0$, for all $l$, to ensure that $\Lambda_{0}(\cdot)$ is monotone increasing function. Briefly, the spline basis functions are piecewise polynomial functions and are fully determined once a knot sequence and the degree are specified. The shape of the basis splines are predominantly determined by the placement of the knots while the degree controls the smoothness \citep{cai2011}. For instance, specifying the degree to take values 1, 2 or 3 correspond to the use of linear, quadratic or cubic polynomials respectively.  Given the $m$ knots and degree, the $k$ ($k=m +$ degree - 2) basis functions are fully determined. For further discussion on specifying the knots, as well as their placement, see \cite{wang2016}, \cite{ramsay}, and \cite{McMahanWangTebbs2013}.

\subsection{Data Augmentation}
Under either of the representations of  $\Lambda_{0}(\cdot)$ proposed in Section 2.2, the unknown parameters in the observed data likelihood consist of $\boldsymbol{\theta}=(\boldsymbol{\beta}', \boldsymbol{\gamma}',\alpha)'$, where $\boldsymbol{\gamma}=(\gamma_{1},\ldots, \gamma_{k})'$. Since the observed data likelihood exists in closed-form, the maximum likelihood estimator MLE of $\boldsymbol{\theta}$ could be obtained by directly maximizing \eqref{eqn:odl1} with respect to $\boldsymbol{\theta}$; i.e., one could obtain $\hat{\boldsymbol{\theta}}$, the MLE of $\boldsymbol{\theta}$, as $\hat{\boldsymbol{\theta}}=\mathrm{argmax}_{\boldsymbol{\theta}}L_{obs}(\boldsymbol{\theta})$. It is worthwhile to point out that the numerical process of directly maximizing \eqref{eqn:odl1}, with respect to $\boldsymbol{\theta}$, is often unstable and rarely performs well \citep{WangWangMcMahan}.

To circumvent these numerical instabilities, an EM algorithm was derived for the purposes of identifying the MLE. This algorithm was developed based on a two-stage data augmentation process, where carefully structured latent Poisson random variables are introduced as missing data. The first stage relates both the instantaneous failure indicator and the censoring indicators to latent Poisson random variables; i.e., the $Z_{i}$, $W_{i}$, and $Y_{i}$ are introduced such that 
\begin{eqnarray*}
Z_{i} &\sim&\textrm{Poisson}\{\Lambda_{0}(t_{i1})\exp({\bf x}_i'\boldsymbol{\beta})\}, \\
W_{i} &\sim& \textrm{Poisson}[\{\Lambda_{0}(t_{i2})-\Lambda_{0}(t_{i1})\}\exp({\bf x}_i'\boldsymbol{\beta})],\\
Y_{i} &\sim&\textrm{Poisson}\{\alpha\exp({\bf x}_i'\boldsymbol{\beta})\},
\end{eqnarray*}
subject to the following constraints: $\delta_{i1}=I(Z_{i}>0)$, $\delta_{i2}=I(Z_{i}=0,W_{i}>0)$, $\delta_{i3}=I(Z_{i}=0,W_{i}=0)$, and $\psi_i=I(Y_i=0)$ where $t_{i1}=R_{i} I(\delta_{i1}=1)+L_{i} I(\delta_{i1}=0)$, and
$t_{i2}=R_{i} I(\delta_{i2}=1)+L_{i} I(\delta_{i3}=1)$. At this stage of the data augmentation, the conditional likelihood is 
\begin{equation}\label{eqn:la}
L_A(\boldsymbol{\theta})=\prod_{i=1}^{n}\Big\{P_{Z_{i}}(Z_{i})P_{W_{i}}(W_{i})^{\delta_{i2}+\delta_{i3}}C_{i}\Big\}^{\psi_i} P_{Y_{i}}(Y_{i})I(Y_{i}=0)^{\psi_i}I(Y_{i}>0)^{(1-\psi_i)},
\end{equation}
where $C_{i}=\delta_{i1}I(Z_{i}>0)+ \delta_{i2}I(Z_{i}=0,W_{i}>0)+\delta_{i3}I(Z_{i}=0,W_{i}=0)$ and $P_A(\cdot)$ is the probability mass function of the random variable $A$. 
In the second and final stage, the $Z_{i}$ and $W_{i}$ are separately decomposed into the sum of $k$ independent latent Poisson random variables; i.e., $Z_{i}=\sum_{l=1}^{k}Z_{il}$ and $W_{i}=\sum_{l=1}^{k}W_{il}$, where 
\begin{eqnarray*}
Z_{il} &\sim& \textrm{Poisson}\{\gamma_{l}b_{l}(t_{i1})\exp({\bf x}_i'\boldsymbol{\beta})\} \textrm{ and } \\
W_{il} &\sim& \textrm{Poisson}[\{\gamma_{l}b_{l}(t_{i2})-\gamma_{l}b_{l}(t_{i1})\}\exp({\bf x}_i'\boldsymbol{\beta})].
\end{eqnarray*}
At this stage of the augmented data likelihood is 
\begin{eqnarray}\label{eqn:lc}
L_C(\boldsymbol{\theta})=\prod_{i=1}^{n}\prod_{l=1}^{k}\Big[P_{Z_{il}}(Z_{il}) I(Z_{i}=Z_{i\cdot})  \{P_{W_{il}}(W_{il}) I(W_{i}=W_{i\cdot})\}^{\delta_{i2}+\delta_{i3}}C_{i}\Big]^{\psi_i} P_{Y_{i}}(Y_{i})I(Y_{i}=0)^{\psi_i}I(Y_{i}>0)^{(1-\psi_i)},
\end{eqnarray} 
where $Z_{i\cdot}=\sum_{l=1}^k Z_{il}$ and $W_{i\cdot}=\sum_{l=1}^k W_{il}$. It is relatively easy to show that by integrating \eqref{eqn:lc} over the latent random variables one will obtain the observed data likelihood depicted in \eqref{eqn:odl1}.

\subsection{EM algorithm}
In general, the EM algorithm consists of two steps: the expectation step (E-step) and the maximization step (M-step). The E-step in this algorithm involves taking the expectation of $\log\{L_c(\boldsymbol{\theta})\}$ with respect to all latent variables conditional on the current parameter value $\boldsymbol{\theta}^{(d)}=(\boldsymbol{\beta}^{(d)'},\boldsymbol{\gamma}^{(d)'},\alpha^{(d)})'$ and the observed data $\boldsymbol{\mathcal{D}}$. This results in obtaining the $Q(\boldsymbol{\theta},\boldsymbol{\theta}^{(d)})$ function, where $Q(\boldsymbol{\theta},\boldsymbol{\theta}^{(d)})=E[\log\{L_c(\boldsymbol{\theta})\}|\boldsymbol{\mathcal{D}},\boldsymbol{\theta}^{(d)}]$. The M-step then finds $\boldsymbol{\theta}^{(d+1)}=\mathrm{argmax}_{\boldsymbol{\theta}} Q(\boldsymbol{\theta},\boldsymbol{\theta}^{(d)})$. This process is repeated in turn until convergence of the algorithm is attained. In this particular setting, the E-step yields $Q(\boldsymbol{\theta},\boldsymbol{\theta}^{(d)})$ as
\begin{eqnarray*}
Q(\boldsymbol{\theta},\boldsymbol{\theta}^{(d)})&=&\sum_{i=1}^{n}\sum_{l=1}^{k}\psi_i\Big[\{E(Z_{il})+(\delta_{i2}+\delta_{i3})E(W_{il})\}\{\log(\gamma_{l})+{\bf x}_i'\boldsymbol{\beta}\}
-\gamma_{l} e^{\mathbf{x}_i'\boldsymbol{\beta}}\{(\delta_{i2}+\delta_{i1})b_{l}(R_i)+\delta_{i3}b_{l}(L_i)\}\Big]\\
&& +\sum_{i=1}^{n} E(Y_i)\log(\alpha e^{{\bf x}_i'\boldsymbol{\beta}})-\alpha e^{{\bf x}_i'\boldsymbol{\beta}}+ H(\boldsymbol{\theta}^{(d)}),
\end{eqnarray*}
where $H(\boldsymbol{\theta}^{(d)})$ is a function of $\boldsymbol{\theta}^{(d)}$ but is free of $\boldsymbol{\theta}$. Notice that in $Q(\boldsymbol{\theta},\boldsymbol{\theta}^{(d)})$ we suppress, for notational convenience, the dependence of the expectations on the observed data and $\boldsymbol{\theta}^{(d)}$; i.e., from henceforth it should be understood that $E(\cdot)=E(\cdot|\boldsymbol{\mathcal{D}},\boldsymbol{\theta}^{(d)})$.

An enticing feature, which makes the proposed approach computationally efficient, is that all of the expectations in $Q(\boldsymbol{\theta},\boldsymbol{\theta}^{(d)})$ can be expressed in closed-form, and moreover can be computed via simple matrix and vector operations. In particular, from \eqref{eqn:la} it can be ascertained that if $\delta_{i1}=1$ and $\psi_i=1$ then $Z_i$ conditionally, given $\boldsymbol{\theta}^{(d)}$ and $\boldsymbol{\mathcal{D}}$, follows a zero-truncated Poisson distribution, and it follows a degenerate distribution at 0 for any other values of $\delta_{i1}$ and $\psi_i$. Thus, the conditional expectation of $Z_i$, given $\boldsymbol{\theta}^{(d)}$ and $\boldsymbol{\mathcal{D}}$, can be expressed as 
\begin{equation*}
E(Z_i)= \delta_{i1}\psi_i \Lambda_0^{(d)}(t_{i1})\exp({\bf x}_i'\boldsymbol{\beta}^{(d)})    \left[1-\exp\{-\Lambda_{0}^{(d)}(t_{i1})\exp({\bf x}_i'\boldsymbol{\beta}^{(d)})\} \right]^{-1},
\end{equation*}
where $\Lambda_{0}^{(d)}(t)=\sum_{l=1}^{k}\gamma_{l}^{(d)}b_{l}(t)$. Through a similar set of arguments one can obtain the necessary conditional expectations of $W_i$ and $Y_i$ as 
\begin{eqnarray*}
E(W_i)&=& \delta_{i2}\psi_i \{\Lambda_0^{(d)}(t_{i2})-\Lambda_0^{(d)}(t_{i1})\}\exp({\bf x}_i'\boldsymbol{\beta}^{(d)})   \left(1-\exp[-\{\Lambda_0^{(d)}(t_{i2})-\Lambda_{0}^{(d)}(t_{i1})\}\exp({\bf x}_i'\boldsymbol{\beta}^{(d)})] \right)^{-1},\\
E(Y_i)&=& (1-\psi_i)\alpha^{(d)}\exp({\bf x}_i'\boldsymbol{\beta}^{(d)})  \left[1-\exp\{-\alpha^{(d)}\exp({\bf x}_i'\boldsymbol{\beta}^{(d)})\}\right]^{-1},
\end{eqnarray*}
respectively. Further, from \eqref{eqn:lc} it can be ascertained that if $\delta_{i1}=1$ and $\psi_i=1$ then $Z_{il}$ conditionally, given $Z_i, \boldsymbol{\mathcal{D}}$ and $\boldsymbol{\theta}^{(d)}$, follows a binomial distribution with $Z_i$ being the number of trials and $\gamma_{l}^{(d)}b_{l}(t_{i1}) \{\Lambda_{0}^{(d)}(t_{i1})\}^{-1}$ being the success probability, and it is follows a degenerate distribution at 0 for any other values of $\delta_{i1}$ and $\psi_i$. Thus, through an application of the Law of Iterated Expectations, the conditional expectation of $Z_{il}$, given $\boldsymbol{\theta}^{(d)}$ and $\boldsymbol{\mathcal{D}}$, can be expressed as
\begin{equation*}
E(Z_{il})= E(Z_i)\gamma_{l}^{(d)}b_{l}(t_{i1}) \{\Lambda_{0}^{(d)}(t_{i1})\}^{-1}.
\end{equation*}
Through a similar set of arguments one can obtain the necessary conditional expectation of $W_{il}$ as 
\begin{equation*}
E(W_{il})=E(W_i)\gamma_{l}^{(d)}\{b_{l}(t_{i2})-b_{l}(t_{i1})\} \{\Lambda_{0}^{(d)}(t_{i2})-\Lambda_{0}^{(d)}(t_{i1})\}^{-1}.
\end{equation*}
Note, in the expressions of the expectations of $Z_{il}$ and $W_{il}$ the dependence on $\delta_{i1}$, $\delta_{i2}$, and $\psi_i$ are suppressed with the properties associated with these variables being inherited from the expectations associated with $Z_i$ and $W_i$, respectively.

The M-step of the algorithm then finds $\boldsymbol{\theta}^{(d+1)}=\mathrm{argmax}_{\boldsymbol{\theta}} Q(\boldsymbol{\theta},\boldsymbol{\theta}^{(d)})$. To this end, consider the partial derivatives of $Q(\boldsymbol{\theta},\boldsymbol{\theta}^{(d)})$ with respect to $\boldsymbol{\theta}$ which are given by

\begin{eqnarray}
\frac{\partial Q(\boldsymbol{\theta},\boldsymbol{\theta}^{(d)})}{\partial\gamma_{l}}&=&\sum_{i=1}^{n}\psi_i\big[\gamma_{l}^{-1}\{E(Z_{il})+(\delta_{i2}+\delta_{i3})E(W_{il})\}- e^{\mathbf{x}_i'\boldsymbol{\beta}}\{(\delta_{i2}+\delta_{i1})b_l(R_i)+\delta_{i3}b_l(L_i)\} \big],\label{eqn:partialgamma}\\
\frac{\partial Q(\boldsymbol{\theta},\boldsymbol{\theta}^{(d)})}{\partial\alpha}&=&\sum_{i=1}^{n} -e^{{\bf x}_i'\boldsymbol{\beta}}+\alpha^{-1}E(Y_i)\label{eqn:partialalpha},\\
\frac{\partial Q(\boldsymbol{\theta},\boldsymbol{\theta}^{(d)})}{\partial\boldsymbol{\beta}}&=&\sum_{i=1}^{n}\big[ \psi_i\{E(Z_{i})+\delta_{i2}E(W_{i})\}-\psi_i\{(\delta_{i1}+\delta_{i2})\Lambda_{0}(R_i)+\delta_{i3}\Lambda_{0}(L_i)\}e^{\mathbf{x}_i'\boldsymbol{\beta}} -\alpha e^{{\bf x}_i'\boldsymbol{\beta}}+E(Y_i) \big]{\bf x}_i. ~~~ \label{eqn:partialbeta}
\end{eqnarray}

By setting \eqref{eqn:partialgamma} equal to zero and solving for $\gamma_{l}$, one can obtain
\begin{equation}\label{eqn:g}
\gamma_l^{*}(\boldsymbol{\beta})=\frac{\sum_{i=1}^{n}\psi_i\{E(Z_{il})+\delta_{i2}E(W_{il})\}}{\sum_{i=1}^{n}\psi_i\{(\delta_{i2}+\delta_{i1})b_l(R_i)+\delta_{i3}b_l(L_i)\}e^{{\bf x}_i'\boldsymbol{\beta}}}, 
\end{equation}
for $l=1,\dots,k$. Similarly, by setting \eqref{eqn:partialalpha} equal to zero and solving for $\alpha$, one can obtain
\begin{equation}\label{eqn:a}
\alpha^{*}(\boldsymbol{\beta})=\frac{\sum_{i=1}^{n}E(Y_i)}{\sum_{i=1}^{n}e^{{\bf x}_i'\boldsymbol{\beta}}}.
\end{equation}
Notice that, $\gamma_l^{*}(\boldsymbol{\beta})$ and $\alpha^{*}(\boldsymbol{\beta})$ depend on $\boldsymbol{\beta}$. Thus, one can obtain $\boldsymbol{\beta}^{(d+1)}$ by setting \eqref{eqn:partialbeta} equal to zero and solving the resulting system of equations for $\boldsymbol{\beta}$, after replacing $\gamma_l$ and $\alpha$ by $\gamma_l^{*}(\boldsymbol{\beta})$ and $\alpha^{*}(\boldsymbol{\beta})$, respectively. Note, the aforementioned system of equations can easily be solved using a standard Newton Raphson approach. Finally, one obtains $\gamma_{l}^{(d+1)}$ and $\alpha^{(d+1)}$ as $\gamma_l^{*}(\boldsymbol{\beta}^{(d+1)})$ and $\alpha^{*}(\boldsymbol{\beta}^{(d+1)})$, respectively.

The proposed EM algorithm is now succinctly stated. First, initialize $\boldsymbol{\theta}^{(0)}$ and repeat the following steps until converges.

\begin{enumerate}
	\item Calculate $\boldsymbol{\beta}^{(d+1)}$ by solving the following system of equations 
	\begin{eqnarray*}
         \sum_{i=1}^{n}\big[\psi_i \{E(Z_{i})+\delta_{i2}E(W_{i})\}  -\alpha^{*}(\boldsymbol{\beta}) e^{{\bf x}_i'\boldsymbol{\beta}}+E(Y_i)  \big]{\bf x}_i = \sum_{i=1}^{n}\sum_{l=1}^{k}\psi_i\{(\delta_{i1}+\delta_{i2})b_l(R_i)+\delta_{i3}b_l(L_i)\}\gamma_l^{*}(\boldsymbol{\beta})e^{{\bf x}_i'\boldsymbol{\beta}}     {\bf x}_i,\\
	\end{eqnarray*}
	where $\gamma_l^{*}(\boldsymbol{\beta})$ and $\alpha^{*}(\boldsymbol{\beta})$ are defined above. 
	\item Calculate $\gamma_{l}^{(d+1)}=\gamma_{l}^{*}\big(\boldsymbol{\beta}^{(d+1)}\big)$ for  $l=1,\dots,k$ and $\alpha^{(d+1)}=\alpha^{*}\big(\boldsymbol{\beta}^{(d+1)}\big)$ .
	\item Update $d=d+1$.	
\end{enumerate}
At the point of convergence, define $\boldsymbol{\theta}^{(d)} =(\boldsymbol{\beta}^{(d)'},\boldsymbol{\gamma}^{(d)'},\alpha^{(d)})'$ to be the proposed estimator $\hat{\boldsymbol{\theta}}=(\hat{\boldsymbol{\beta}}',\hat{\boldsymbol{\gamma}}',\hat{\alpha})'$, which is the MLE of $\boldsymbol{\theta}$.

\subsection{Variance estimation}

For the purposes of uncertainty quantification, several variance estimators were considered and evaluated; e.g., the inverse of the observed Fisher information matrix, the Huber sandwich estimator, and the outer product of gradients (OPG) estimator. After extensive numerical studies (results not shown), it was found that the most reliable variance estimator, among those considered, was that of the OPG estimator. In general, the OPG estimator is obtained as
\begin{equation*}
\widehat{V}(\widehat{\boldsymbol{\theta}})=\left[\frac{1}{n}\sum_{i=1}^{n}\dot{l}_i(\widehat{\boldsymbol{\theta}})\dot{l}_i^{'}(\widehat{\boldsymbol{\theta}})\right]^{-1}
\end{equation*}
where $\dot{l}_i(\widehat{\boldsymbol{\theta}})=\partial l_i(\boldsymbol{\theta})/\partial \boldsymbol{\theta} |_{\boldsymbol{\theta}=\widehat{\boldsymbol{\theta}}}$ and $l_i(\boldsymbol{\theta})$ is the log-likelihood contribution of the $i$th individual. Using this estimator, one can conduct standard Wald type inference.

\section{Simulation Study}
In order to investigate the finite sample performance of the proposed methodology, the following simulation study was conducted. The true distribution of the failure times was specified to be
\[
   H(t|\mathbf{x})=\left\{
                \begin{array}{ll}
                  1-e^{-\alpha e^{\mathbf{x}'\boldsymbol{\beta}}}, &\textrm{ for } t=0,\\
                  1-e^{-\alpha e^{\mathbf{x}'\boldsymbol{\beta}}}\{1-F(t|\mathbf{x})\}, &\textrm{ for } t>0,
                 
                \end{array}
              \right.
  \]
where $p=0.3$ (i.e., $\alpha= -\log(0.7)$), $\mathbf{x}=({\bf x}_1, {\bf x}_2)'$, ${\bf x}_1 \sim \textrm{N}(0,1)$, ${\bf x}_2\sim \textrm{Bernoulli}(0.5)$, and $\boldsymbol{\beta}=(\beta_1, \beta_2)'$, where $\beta_{1}$ and $\beta_{2}$ take on values of -0.5 and 0.5 resulting 4 regression parameter configurations. Additionally, these studies consider two cumulative baseline hazard functions; i.e., a logarithmic $\Lambda_{0}(t)=\log(t+1)/\log(11)$ and a linear $\Lambda_{0}(t)=0.1t$. These choices were made so that the hazard functions behave similarly but have different shapes. In total, these specifications lead to eight separate data generating models for the failure times. Two generating processes were considered for the observation times: an exponential distribution with a mean of 10 and a discrete uniform over the interval $[1,17]$. In both cases, a single observation time, $O$, was generated for each failure time which was not instantaneous (i.e., $T>0$), and the intervals were created such that $L=0$ ($R=\infty$) and $R=O$ ($L=O$) if $T$ was smaller (greater) than $O$. A few comments are warranted on the selection of the observation processes. First, the former process attempts to match the baseline characteristics of the failure time distribution. Second, note that the specification of the two observation processes result in case-1 interval-censored (i.e., current status) data. This was done for a primary reason that the data of this nature possess less information when compared to general interval-censored data, and thus if the proposed approach works well in this setting it could be inferred that it should perform better in the case of interval-censored data. In total, these data generating steps lead to sixteen generating mechanisms, and each were used to create 500 independent data sets consisting of $n=100$ observations.          

In order to examine the performance of the proposed approach across a broad spectrum of characteristics, several different models were used to analyze each data set. First, following the development presented in Section 2.2, three different parametric forms were considered for the cumulative baseline hazard function: $\Lambda_{0_1}(t)=\gamma_1\log(t+1)$, $\Lambda_{0_2}(t)=\gamma_1t$, and $\Lambda_{0_3}(t)=\gamma_1t+\gamma_2t^2$, which are henceforth referred to as models M1, M2, and M3, respectively. Note, these specifications allow one to examine the performance of the proposed approach when the cumulative baseline hazard function is correctly specified (e.g., M2 when $\Lambda_{0}(t)=0.1t$), over specified (e.g., M3 when $\Lambda_{0}(t)=0.1t$), and misspecified (e.g., M1 when $\Lambda_{0}(t)=0.1t$). Further, for each data set a model (M4) was fit using the monotone spline representation for the cumulative baseline hazard function developed in Section 2.2. In order to specify the basis functions, the degree was specified to be 2, in order to provide adequate smoothness, and one interior knot was placed at the median of the observed finite nonzero interval end points, with boundary knots being placed at the minimum and maximum of the same. The EM algorithm derived in Section 2.4 was used to complete model fitting for M1-M4. The starting value for all implementations was set to be $\boldsymbol{\theta}^{(0)}=(\boldsymbol{\beta}^{(0)'},\boldsymbol{\gamma}^{(0)'}, \alpha^{(0)})=(\textbf{0}_2',\textbf{1}_k',0.1)$, where $\textbf{0}_k$($\textbf{1}_k$) is a $(k\times 1)$-dimensional vector of zeros (ones). Convergence was declared when the maximum absolute differences between the parameter updates were less than the specified tolerance of $1\times 10^{-5}$.

Table 1 summarizes the estimates of the regression coefficients and the baseline instantaneous failure probability for all considered simulation configurations and models, when the observation times were drawn from a exponential distribution. This summary includes the empirical bias, the sample standard deviation of the 500 point estimates, the average of the 500 standard error estimates, and the empirical coverage probabilities associated with 95\% Wald confidence intervals. Table 2 provides the analogous results for the case in which the observation times are sampled from a discrete uniform distribution. From these results, one will first note that across all considered simulation settings the proposed approach performs very well for M4 and the correct parametric model (i.e., M1 when $\Lambda_{0}(t)=\log(t+1)/\log(11)$ and M2 when $\Lambda_{0}(t)=0.1t$); i.e., the parameter estimates exhibit very little bias, the sample standard deviation of the 500 point estimates are in agreement with the average of the standard error estimates, and the empirical coverage probabilities are at their nominal level. In summary, these findings tend to suggest that the proposed methodology can be used to reliably estimate the covariate effects, the instantaneous failure probability, and quantify the uncertainty in the same. Additionally, these findings generally continue to persist for the case in which the parametric model is over specified (e.g., M3 when $\Lambda_{0}(t)=0.1t$), with the resulting estimates in some cases exhibiting a slightly larger bias and a bit more variability, as one would expect. Further, from these results one will also note that when the parametric model is misspecified (e.g., M2 and M3 when $\Lambda_{0}(t)=\log(t+1)/\log(11)$) the estimates tend to exhibit more bias and less reliable inference, which is expected under the misspecification of the cumulative baseline hazard function. Finally, the estimates obtained under M4 (i.e., the model which makes use of the monotone splines) exhibit little if any difference when compared to the estimates resulting from correct parametric model. In totality, from these findings, it is conjectured that the approach which makes use of the spline representation to approximate the unknown cumulative baseline hazard functions (i.e., M4) would likely be preferable, since it avoids the potential of model misspecification and it obtains estimators of the unknown parameters, as well as their standard errors, that are equivalent to those estimators obtained under the true parametric model, the form of which is generally not known.

\renewcommand{\baselinestretch}{1}
\begin{table}\tiny
\begin{center}
\begin{tabular}{lrrrrcrrrrcrrrrcrrrr}
\hline\hline
&&&&&&&&&&&&&&&&&&& \\ [-3pt]
&\multicolumn{16}{c}{True $\Lambda_0(t)=\log(t+1)/\log(11)$}\\ [2pt]\hline
&&&&&&&&&&&&&&&&&&& \\ [-3pt]
&\multicolumn{4}{c}{M1(True)} &&\multicolumn{4}{c}{M2(Misspecified)}&&\multicolumn{4}{c}{M3(Misspecified)}&&\multicolumn{4}{c}{M4(Spline)}\\ [2pt]
\cline{2-5} \cline{7-10} \cline{12-15}\cline{17-20}
&&&&&&&&&&&&&&&&&&&\\ [-3pt]
Parameter&Bias&SD&ESE&CP95&&Bias&SD&ESE&CP95&&Bias&SD&ESE&CP95&&Bias&SD&ESE&CP95\\ [2pt]  \hline\\[-3pt]
$\beta_{1}=-0.5$&  -0.03 & 0.15 &0.15  & 0.94 && -0.06 & 0.17 &0.15 & 0.89&& -0.06 & 0.17 &0.15 &0.90 && -0.04 & 0.16 &  0.16& 0.93 \\ 
$\beta_{2}=-0.5$&  -0.01 & 0.29 &0.28  & 0.94 &&-0.05  & 0.33 & 0.27&0.89 && -0.05 &  0.33& 0.28&0.90 && -0.02 & 0.30 & 0.29 & 0.93 \\ 
$p=0.3$& 0.00  & 0.06 & 0.06 & 0.95 && 0.00 &0.06  &0.06 &0.94 && 0.00 & 0.06 &0.06 &0.94 && 0.00 & 0.06 &0.06  & 0.95 \\ 
&&&&&&&&&&&&&&&&&&&\\ [-3pt]\hline\\[-3pt]
$\beta_{1}=-0.5$&  -0.03 & 0.15 & 0.14 & 0.94 && -0.07 & 0.17 & 0.14& 0.87&& -0.07 & 0.17 & 0.14&0.88 && -0.04 &0.15  &0.15  & 0.94 \\ 
$\beta_{2}=0.5$&  0.01 & 0.26 & 0.26 &0.94 && 0.06 & 0.30 &0.25 & 0.89&& 0.06 & 0.30 &0.26 &0.89 && 0.03 & 0.27 & 0.27 &0.94  \\ 
$p=0.3$&  0.00 & 0.06 & 0.06 & 0.94 && -0.01 & 0.06 & 0.05& 0.92&&-0.01  & 0.06 & 0.06&0.93 && -0.01 & 0.06 & 0.06 & 0.94 \\ 
&&&&&&&&&&&&&&&&&&&\\ [-3pt]\hline\\[-3pt]
$\beta_{1}=0.5$&  0.01 & 0.15 & 0.15 & 0.95 && 0.05 & 0.16 &0.15 & 0.91&& 0.05 & 0.16 &0.15 &0.92 && 0.02 & 0.15 &0.15  & 0.95 \\ 
$\beta_{2}=-0.5$&  0.01 & 0.28 & 0.28 & 0.94 &&-0.03  &0.32  &0.27 &0.90 && -0.03 &0.32 &0.27 &0.91 && 0.01 & 0.29 &0.29  & 0.94 \\ 
$p=0.3$&  0.00 & 0.05 &0.06  & 0.97 && 0.00 & 0.06 & 0.06&0.96 && 0.00 & 0.06 &0.06 &0.96 && 0.00 & 0.05 & 0.06 & 0.96 \\ 
&&&&&&&&&&&&&&&&&&&\\ [-3pt]\hline\\[-3pt]
$\beta_{1}=0.5$& 0.02 & 0.14 & 0.14 & 0.95 && 0.07 &  0.15&0.14 &0.89 && 0.07 & 0.15 &0.14 &0.90 && 0.03  &  0.14& 0.15 &0.95  \\ 
$\beta_{2}=0.5$&  0.02 & 0.28 & 0.26 & 0.94 && 0.06 & 0.32 &0.25 & 0.86&&0.06  &  0.32&0.26 &0.87 && 0.03 & 0.28 &0.27  & 0.93 \\ 
$p=0.3$&  0.00 & 0.06 & 0.06 & 0.93 && -0.01& 0.06 &0.05 &0.90 && -0.01 &  0.06&0.06 &0.91 && 0.00 & 0.06 & 0.06 & 0.93 \\ 
&&&&&&&&&&&&&&&&&&&\\ [-3pt]\hline\hline\\[-3pt]
&\multicolumn{16}{c}{True $\Lambda_0(t)=0.1t$}\\ [2pt]\hline
&&&&&&&&&&&&&&&&&&& \\ [-3pt]
&\multicolumn{4}{c}{M1(Misspecified)} &&\multicolumn{4}{c}{M2(True)}&&\multicolumn{4}{c}{M3(Over specified)}&&\multicolumn{4}{c}{M4(Spline)}\\ [2pt]
\cline{2-5} \cline{7-10} \cline{12-15}\cline{17-20}
&&&&&&&&&&&&&&&&&&&\\ [-3pt]
Parameter&Bias&SD&ESE&CP95&&Bias&SD&ESE&CP95&&Bias&SD&ESE&CP95&&Bias&SD&ESE&CP95\\ [2pt]  \hline\\[-3pt]
$\beta_{1}=-0.5$&  0.01 & 0.15 & 0.16 & 0.96 && -0.03 & 0.16 &0.16 & 0.94&& -0.05 & 0.17 & 0.16&0.94 && -0.04 & 0.16 & 0.17 & 0.94 \\ 
$\beta_{2}=-0.5$& 0.04 & 0.28 & 0.29& 0.95 && 0.00 & 0.30 & 0.30& 0.95&& -0.03 & 0.32 &0.30 &0.94 && -0.02  & 0.32 & 0.31 &0.95  \\ 
$p=0.3$& 0.00 & 0.06 & 0.06 & 0.95 && 0.00 & 0.06 &0.06 & 0.95&& 0.00 & 0.06 & 0.06&0.95 && 0.00  & 0.06 & 0.06 & 0.95 \\ 
&&&&&&&&&&&&&&&&&&&\\ [-3pt]\hline\\[-3pt]
$\beta_{1}=-0.5$&  0.02 & 0.15 & 0.15 & 0.94 && -0.02 & 0.15 &0.15 & 0.94&& -0.04 &  0.16&0.15 &0.93 && -0.04 & 0.16 & 0.16 & 0.94 \\ 
$\beta_{2}=0.5$&  -0.02 &0.26  & 0.27 & 0.96 &&  0.02& 0.28 & 0.27&0.93 && 0.03 & 0.29 &0.28 &0.93 && 0.03 & 0.29 & 0.29 & 0.95 \\ 
$p=0.3$&  0.01 & 0.06 & 0.06 & 0.94 && 0.00 & 0.06 & 0.06& 0.94&& -0.01 &0.06  & 0.06&0.94 && -0.01 & 0.06 & 0.06 &  0.94\\ 
&&&&&&&&&&&&&&&&&&&\\ [-3pt]\hline\\[-3pt]
$\beta_{1}=0.5$& -0.02 & 0.14 & 0.16 & 0.96 && 0.02 & 0.15 & 0.16&0.96 && 0.03 & 0.16 & 0.16&0.95 && 0.03  & 0.16 & 0.17 & 0.95 \\ 
$\beta_{2}=-0.5$&  0.06 & 0.29 & 0.29 & 0.94 && 0.02 & 0.30 &0.29 &0.93 &&0.01  & 0.31 &0.30 &0.93 && 0.01 & 0.31 & 0.31 & 0.94 \\ 
$p=0.3$&  0.00 & 0.05 & 0.06 & 0.97 && 0.00 & 0.05 & 0.06& 0.96&& 0.00 & 0.05 &0.06 &0.96 && 0.00 & 0.06 & 0.06 &0.97  \\ 
&&&&&&&&&&&&&&&&&&&\\ [-3pt]\hline\\[-3pt]
$\beta_{1}=0.5$& -0.02 & 0.14 & 0.15 & 0.95 && 0.02 & 0.15 & 0.15& 0.95&& 0.04 & 0.15 &0.15 &0.95 && 0.04  & 0.15 & 0.16 & 0.95 \\ 
$\beta_{2}=0.5$&  -0.02 & 0.27 &0.27  & 0.94 && 0.02 & 0.29 &0.27 &0.94 && 0.04 & 0.30 &0.28 &0.93 && 0.04 &0.30  & 0.29 & 0.93 \\ 
$p=0.3$&  0.01 & 0.06 & 0.06 & 0.96 && 0.00 & 0.06 &0.06 &0.95 && -0.01 &0.06  &0.06 &0.94 && -0.01 & 0.06 & 0.06 & 0.95 \\ 
&&&&&&&&&&&&&&&&&&&\\ [-3pt]\hline\hline\\[-3pt]
\end{tabular}
\caption{Simulation results summarizing the estimates of the regression coefficients and the baseline instantaneous failure probability obtained from M1-M4 across all simulation settings, when observation times were sampled from an exponential distribution. This summary include the average of the 500 point estimates minus the true value (Bias), the sample standard deviation of the 500 point estimates (SD), the average of the estimated standard errors (ESE), and empirical coverage probabilities associated with 95\% Wald confidence intervals (CP95). Note, when $\Lambda_0(t)=\log(t+1)/\log(11)$ then M1 is the true parametric model and when $\Lambda_0(t)=0.1t$ then M2 is the true parametric model.}
\label{Ob is exp(0.1)}
\end{center}
\end{table}

\renewcommand{\baselinestretch}{1}
\begin{table}\tiny
\begin{center}
\begin{tabular}{lrrrrcrrrrcrrrrcrrrr}
\hline\hline
&&&&&&&&&&&&&&&&&&& \\ [-3pt]
&\multicolumn{16}{c}{True $\Lambda_0(t)=\log(t+1)/\log(11)$}\\ [2pt]\hline
&&&&&&&&&&&&&&&&&&& \\ [-3pt]
&\multicolumn{4}{c}{M1(True)} &&\multicolumn{4}{c}{M2(Misspecified)}&&\multicolumn{4}{c}{M3(Misspecified)}&&\multicolumn{4}{c}{M4(Spline)}\\ [2pt]
\cline{2-5} \cline{7-10} \cline{12-15}\cline{17-20}
&&&&&&&&&&&&&&&&&&&\\ [-3pt]
Parameter&Bias&SD&ESE&CP95&&Bias&SD&ESE&CP95&&Bias&SD&ESE&CP95&&Bias&SD&ESE&CP95\\ [2pt]  \hline\\[-3pt]
$\beta_{1}=-0.5$& -0.02 & 0.14 & 0.15 & 0.95 && -0.03 & 0.15 & 0.15&0.92 &&-0.03  & 0.15 & 0.15&0.93 && -0.03 & 0.15 &0.15  & 0.94 \\ 
$\beta_{2}=-0.5$&  0.00 & 0.27 & 0.27 & 0.96 &&-0.02  & 0.29 &0.27 &0.94 && -0.02 & 0.29 & 0.27&0.94 && -0.02 & 0.28 & 0.28 & 0.96 \\ 
$p=0.3$&  -0.01 &0.06  & 0.06 &0.95  && -0.01 & 0.06 &0.06 & 0.93&& -0.01 & 0.06 &0.06 &0.94 && -0.01 & 0.06 & 0.06 & 0.94 \\ 
&&&&&&&&&&&&&&&&&&&\\ [-3pt]\hline\\[-3pt]
$\beta_{1}=-0.5$&-0.03  & 0.14 & 0.14 & 0.94 && -0.05 & 0.15 & 0.14& 0.92&&-0.05  & 0.15 &0.14 &0.92 && -0.04 &0.15  & 0.14 & 0.94 \\
$\beta_{2}=0.5$&  0.00 & 0.26 & 0.25 &0.94  && 0.03 &0.28  & 0.25& 0.92&&  0.03& 0.28 &0.25 &0.93 && 0.02& 0.27 & 0.26 &  0.93\\ 
$p=0.3$& 0.00 &  0.06&  0.06&0.95  &&  0.00&0.06  &0.06 &0.94 && 0.00 &0.06  &0.06 &0.94 && 0.00 & 0.06 &0.06  &0.95 \\
&&&&&&&&&&&&&&&&&&&\\ [-3pt]\hline\\[-3pt]
$\beta_{1}=0.5$&0.02  & 0.16 &0.15  & 0.94 && 0.03 & 0.16 & 0.15& 0.94&& 0.03 &0.16  &0.15 &0.94 && 0.03 &0.16  & 0.15 & 0.94 \\ 
$\beta_{2}=-0.5$& 0.00 & 0.28 &0.27  & 0.95 && -0.01 & 0.29 &0.27 & 0.93&& -0.01 & 0.29 &0.27 &0.93 && -0.01 & 0.29 & 0.28 & 0.94 \\
$p=0.3$& 0.00 & 0.06 & 0.06 & 0.95 && 0.00 & 0.06 &0.06 & 0.95&& 0.00 & 0.06 &0.06 &0.95 && 0.00 & 0.06 &0.06  & 0.95 \\
&&&&&&&&&&&&&&&&&&&\\ [-3pt]\hline\\[-3pt]
$\beta_{1}=0.5$& 0.01 & 0.14 &  0.14& 0.95 &&0.04  & 0.15 & 0.14& 0.92&& 0.04 & 0.15 &0.14 &0.93 && 0.03 & 0.14 & 0.14 & 0.94 \\
$\beta_{2}=0.5$& 0.03 & 0.27 & 0.25 &0.94  && 0.05 & 0.29 & 0.25&0.91 && 0.05 & 0.29 & 0.25&0.91 && 0.04 & 0.28 & 0.26 & 0.92 \\
$p=0.3$& 0.00 &0.06  &0.06  & 0.95 && -0.01 &  0.06&0.05 &0.94 && -0.01 & 0.06 &0.06 &0.94 && 0.00&0.06  & 0.06 & 0.94\\
&&&&&&&&&&&&&&&&&&&\\ [-3pt]\hline\hline\\[-3pt]
&\multicolumn{16}{c}{True $\Lambda_0(t)=0.1t$}\\ [2pt]\hline
&&&&&&&&&&&&&&&&&&& \\ [-3pt]
&\multicolumn{4}{c}{M1(Misspecified)} &&\multicolumn{4}{c}{M2(True)}&&\multicolumn{4}{c}{M3(Over specified)}&&\multicolumn{4}{c}{M4(Spline)}\\ [2pt]
\cline{2-5} \cline{7-10} \cline{12-15}\cline{17-20}
&&&&&&&&&&&&&&&&&&&\\ [-3pt]
Parameter&Bias&SD&ESE&CP95&&Bias&SD&ESE&CP95&&Bias&SD&ESE&CP95&&Bias&SD&ESE&CP95\\ [2pt]  \hline\\[-3pt]
$\beta_{1}=-0.5$& 0.00 & 0.15 & 0.15 & 0.95 && -0.02 &0.15  & 0.15& 0.93&&-0.03  &  0.16& 0.16&0.93 && -0.03 & 0.16 & 0.16 & 0.94 \\
$\beta_{2}=-0.5$& 0.02 & 0.27 & 0.28 & 0.96 &&0.00  &  0.28& 0.28&0.96 &&-0.01  & 0.29 & 0.29&0.96 && -0.01 & 0.29 &0.29  & 0.96 \\
$p=0.3$&-0.01  & 0.06 & 0.06 & 0.95 && -0.01 &0.06  &0.06 &0.95 && -0.01 &0.06  &0.06 &0.94 && -0.01 & 0.06 &0.06  & 0.95 \\ 
&&&&&&&&&&&&&&&&&&&\\ [-3pt]\hline\\[-3pt]
$\beta_{1}=-0.5$& 0.00 & 0.14 & 0.14 & 0.95 && -0.02 & 0.14 &0.14 &0.94 && -0.03 & 0.15 & 0.15&0.94 && -0.03 & 0.15 & 0.15 &0.93  \\
$\beta_{2}=0.5$& -0.03  & 0.25 & 0.26 & 0.95 && 0.00 & 0.27 & 0.26& 0.94&& 0.01 &  0.27&0.27 &0.94 && 0.01& 0.27 & 0.27 &0.94  \\
$p=0.3$&  0.01 & 0.06 & 0.06 & 0.96 &&0.00  & 0.06 & 0.06&0.95 && 0.00 & 0.06 &0.06 &0.95 && 0.00&0.06  & 0.06 & 0.95 \\
&&&&&&&&&&&&&&&&&&&\\ [-3pt]\hline\\[-3pt]
$\beta_{1}=0.5$& 0.00 & 0.15 &0.15  & 0.94 && 0.02 & 0.16 &0.15 & 0.94&&  0.02& 0.16 &0.16 &0.94 && 0.02 & 0.16 & 0.16 & 0.94 \\
$\beta_{2}=-0.5$&   0.01& 0.28 & 0.28 & 0.95 && -0.01 & 0.29 &0.28 &0.95 && -0.01 & 0.30 & 0.28&0.94 && -0.01& 0.30 & 0.29 & 0.94 \\
$p=0.3$& 0.00 & 0.06 & 0.06 & 0.96 &&0.00  & 0.06 & 0.06& 0.96&& 0.00 & 0.06 &0.06 &0.96 && 0.00 & 0.06 & 0.06 &0.96  \\
&&&&&&&&&&&&&&&&&&&\\ [-3pt]\hline\\[-3pt]
$\beta_{1}=0.5$&  -0.02 & 0.13 &0.14  &0.96 && 0.01 &0.14  & 0.14& 0.95&& 0.02 &0.14  &0.15 &0.95 && 0.02& 0.14 & 0.15 & 0.96 \\
$\beta_{2}=0.5$&  0.00&0.26  & 0.26 & 0.96 && 0.03 & 0.27 & 0.26& 0.95&& 0.04 & 0.28 & 0.27&0.95 && 0.04 & 0.28 &0.27  & 0.95 \\
$p=0.3$& 0.00  &0.06  &0.06  & 0.96 && 0.00 & 0.06 &0.06 &0.95 && 0.00 & 0.06 & 0.06&0.95 && 0.00& 0.06 & 0.06 & 0.96 \\
&&&&&&&&&&&&&&&&&&&\\ [-3pt]\hline\hline\\[-3pt]
\end{tabular}
\caption{Simulation results summarizing the estimates of the regression coefficients and the baseline instantaneous failure probability obtained from M1-M4 across all simulation settings, when observation times were sampled from a discrete uniform distribution. This summary include the average of the 500 point estimates minus the true value (Bias), the sample standard deviation of the 500 point estimates (SD), the average of the estimated standard errors (ESE), and empirical coverage probabilities associated with 95\% Wald confidence intervals (CP95). Note, when $\Lambda_0(t)=\log(t+1)/\log(11)$ then M1 is the true parametric model and when $\Lambda_0(t)=0.1t$ then M2 is the true parametric model.}
\label{Ob is sample}
\end{center}
\end{table}

Figure 1 summarizes the estimates of the baseline survival function (i.e., $S_{0}(t)=S(t|\mathbf{x}=\mathbf{0}_r)$) obtained from M1-M4 across all considered regression parameter configurations when $\Lambda_0(t)=\log(t+1)/\log(11)$ and the observation times were sampled from the exponential distribution. Figures 2-4 summarizes the analogous results for the other simulation configurations. In particular, these figures present the true baseline survival functions, the average of the point-wise estimates, and the point-wise 2.5th and 97.5th percentiles of the estimates. These figures reinforce the main findings discussed above; i.e., M4 and the correctly specified parametric model again provide reliable estimates of the baseline survival function, and hence the cumulative baseline hazard function, across all simulation configurations. Similarly the over specified model also provides reliable estimates, but the same can not be said for the misspecified models. It is worthwhile to point out that the baseline survival curves do not extend to unity as time goes toward the origin, this is due to the fact that the baseline instantaneous failure probability is $p=0.3$. Again, these findings support the recommendation that the spline based representation of the cumulative baseline hazard function should be adopted in lieu of a parametric model, thus obviating the possible ramifications associated with misspecification.

\begin{figure}
\begin{center}
\includegraphics[width=1\textwidth]{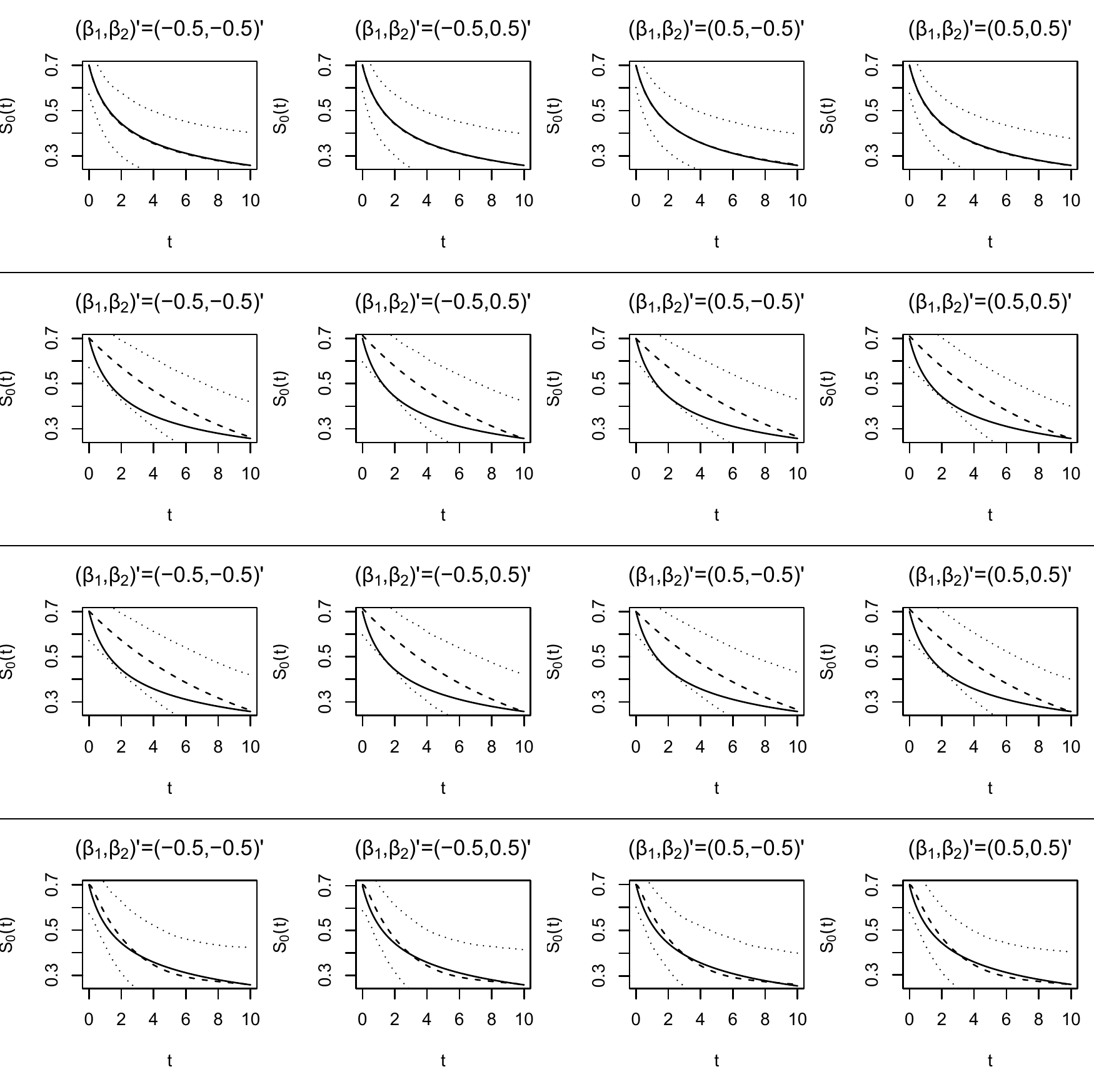}
\caption{Simulation results summarizing the estimates of the baseline survival function obtained by the proposed approach under M1 (first row), M2 (second row), M3 (third row), and M4 (fourth row) when true $\Lambda_0(t)=\log(t+1)/\log(11)$ and observation times were drawn from exponential distribution. The solid line provides the true value, dashed line represents the average estimated value, and the dotted lines indicate the 2.5\% and 97.5\% quantiles, of the point-wise estimates.}
\end{center}
\end{figure}

\begin{figure}
\begin{center}
\includegraphics[width=1\textwidth]{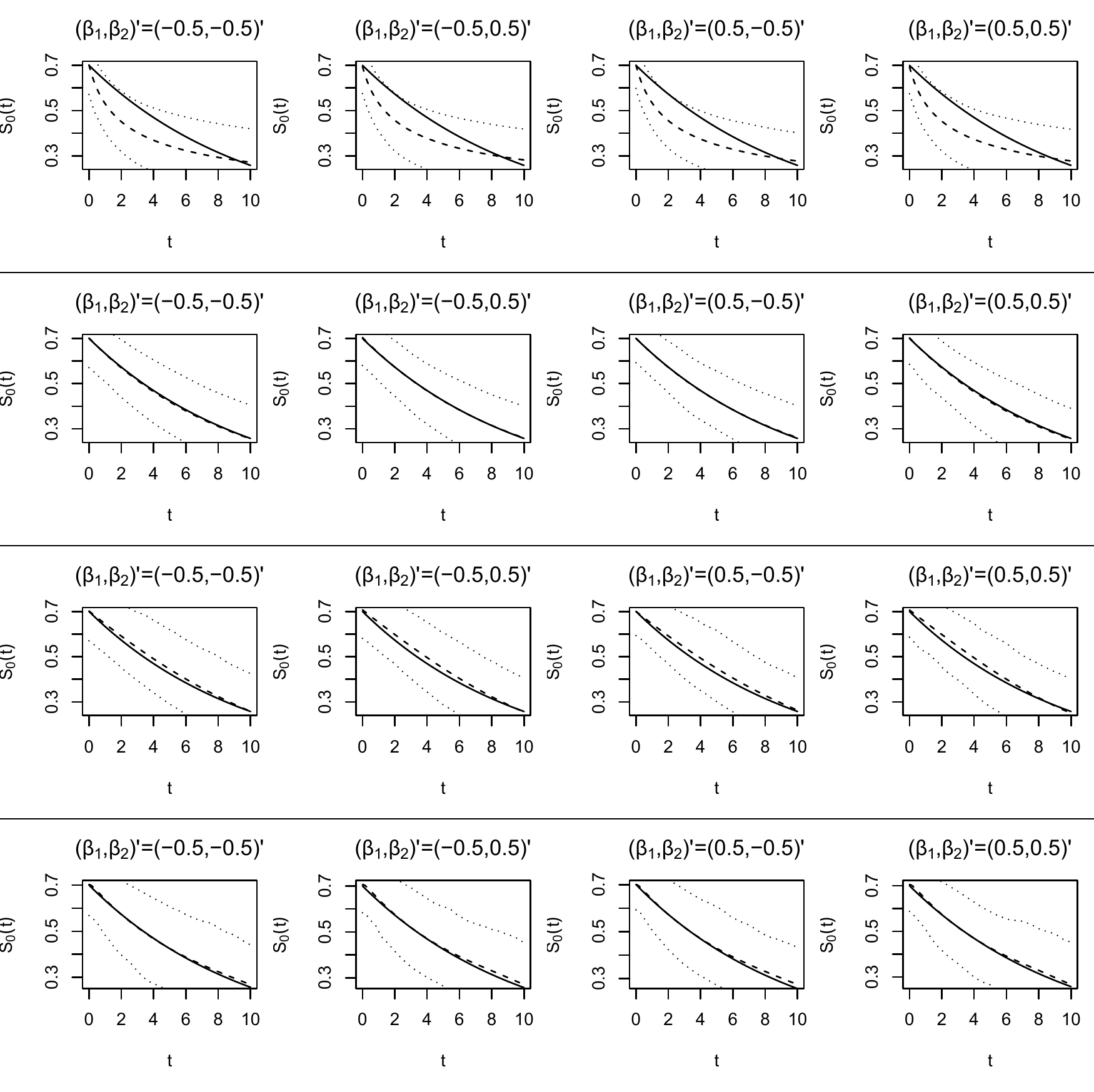}
\caption{Simulation results summarizing the estimates of the baseline survival function obtained by the proposed approach under M1 (first row), M2 (second row), M3 (third row), and M4 (fourth row) when true $\Lambda_0(t)=0.1t$ and observation times were drawn from exponential distribution. The solid line provides the true value, dashed line represents the average estimated value, and the dotted lines indicate the 2.5\% and 97.5\% quantiles, of the point-wise estimates.}
\end{center}
\end{figure}

\begin{figure}
\begin{center}
\includegraphics[width=1\textwidth]{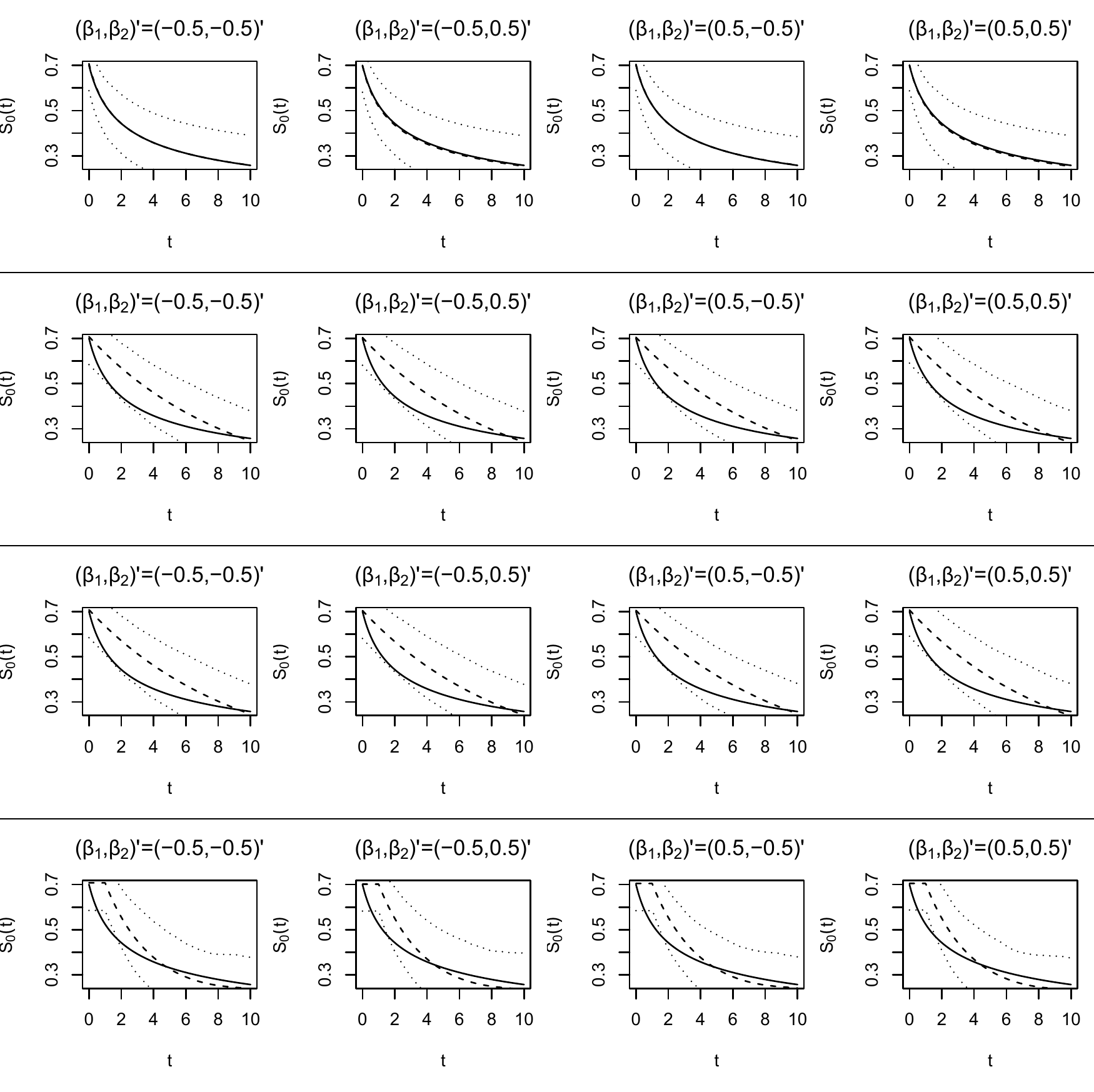}
\caption{Simulation results summarizing the estimates of the baseline survival function obtained by the proposed approach under M1 (first row), M2 (second row), M3 (third row), and M4 (fourth row) when true $\Lambda_0(t)=\log(t+1)/\log(11)$ and observation times were drawn from discrete uniform distribution over [1, 17]. The solid line provides the true value, dashed line represents the average estimated value, and the dotted lines indicate the 2.5\% and 97.5\% quantiles, of the point-wise estimates.}
\end{center}
\end{figure}

\begin{figure}
\begin{center}
\includegraphics[width=1\textwidth]{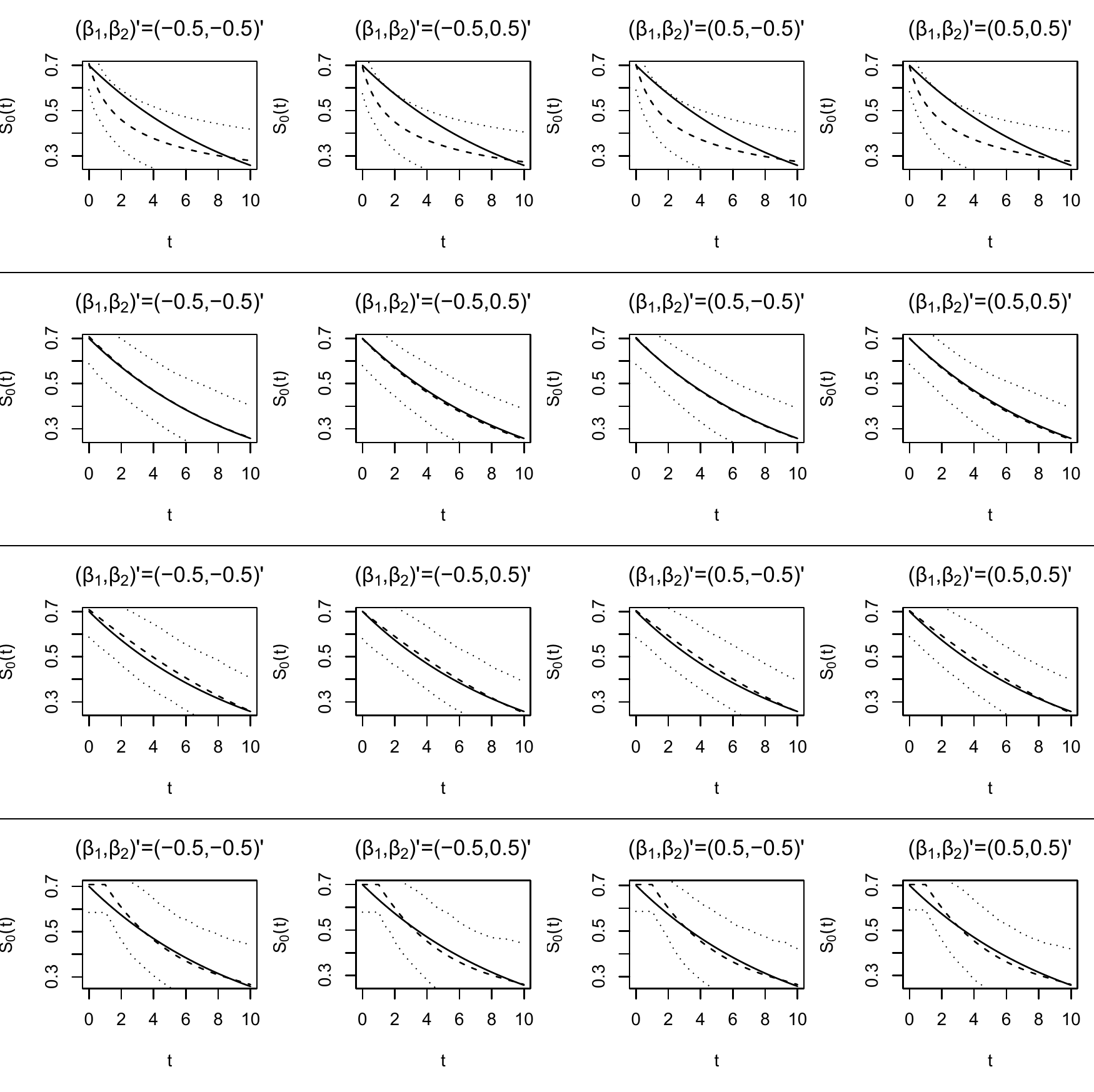}
\caption{Simulation results summarizing the estimates of the baseline survival function obtained by the proposed approach under M1 (first row), M2 (second row), M3 (third row), and M4 (fourth row) when true $\Lambda_0(t)=0.1t$ and observation times were drawn from discrete uniform distribution over [1, 17]. The solid line provides the true value, dashed line represents the average estimated value, and the dotted lines indicate the 2.5\% and 97.5\% quantiles, of the point-wise estimates.}
\end{center}
\end{figure}

In summary, this simulation study illustrates that proposed methodology can be used to analyze current status data which is subject to instantaneous failures, and moreover that the monotone spline approach discussed in Section 2.2 should be adopted for approximating the unknown cumulative baseline hazard function. A few additional details about the numerics of the approach follow. First, the average time required to complete model fitting was approximately one second, supporting the claim that the proposed approach is computationally efficient. Lastly, for the reasons of complete transparency, for a single data set, among 8000, the OPG estimator under M4 resulted in a singular matrix, which prevented standard error estimation estimation, and this issue was resolved by slightly shifting the interior knot.

\section{Discussion}

This work proposed a new model, under the PH assumption, which can be used to analyze interval-censored data subject to instantaneous failures. Through the model development, two techniques for approximating the unknown cumulative baseline hazard function are illustrated. To complete model fitting, a novel EM algorithm is developed through a two-stage data augmentation process. The resulting algorithm is easy to implement and is computationally efficient. These features are likely attributable to the fact that the carefully structured data augmentation steps lead to closed-form expressions for all necessary quantities in the E-step of the algorithm. Moreover, in the M-step the regression coefficients are updated through solving a low-dimensional system of equations, while all other unknown parameters are updated in closed-form. The finite sample performance of the proposed approach was exhibited through an extensive numerical study. This study suggests that the use of the monotone spline representation of the cumulative baseline hazard function would in general be preferable, in order to circumvent the possibility of model misspecification. To further disseminate this work, code, written in R, has been prepared and is available upon request from the corresponding author.

\section{Acknowledgments}
This research was partially supported by National Institutes of Health grant AI121351.


\bibliographystyle{spbasic.bst}   

\bibliography{mybibfile}

%
%

\end{document}